\documentclass[sigconf]{acmart}

\def\BibTeX{{\rm B\kern-.05em{\sc i\kern-.025em b}\kern-.08emT\kern-.1667em\lower.7ex\hbox{E}\kern-.125emX}}
    
\usepackage{url}  
\usepackage{graphicx}  
\usepackage{float}
\usepackage{amsmath}
\usepackage{graphics}
\usepackage[english]{babel}
\usepackage{graphicx}
\usepackage{subcaption}
\begin{document}

%
\title{Decoding the Style and Bias of Song Lyrics}

%

\author{Manash Pratim Barman}
\affiliation{
  \institution{Indian Institute of Information Technology}
  \city{Guwahati}
  \country{India}}

\author{Amit Awekar}
\affiliation{
  \institution{Indian Institute of Technology}
  \city{Guwahati}
  \country{India}}
\email{awekar@iitg.ac.in}

\author{Sambhav Kothari}
\affiliation{
  \institution{Bloomberg LP}
  \city{London}
  \country{United Kingdom}}
\email{skothari44@bloomberg.net}

%

%
\begin{abstract}

The central idea of this paper is to gain a deeper understanding of song lyrics computationally. We focus on two aspects: style and biases of song lyrics. All prior works to understand these two aspects are limited to manual analysis of a small corpus of song lyrics. In contrast, we analyzed more than half a million songs spread over five decades. We characterize the lyrics style in terms of vocabulary, length, repetitiveness, speed, and readability. We have observed that the style of popular songs significantly differs from other songs. We have used distributed representation methods and WEAT test to measure various gender and racial biases in the song lyrics. We have observed that biases in song lyrics correlate with prior results on human subjects. This correlation indicates that song lyrics reflect the biases that exist in society. Increasing consumption of music and the effect of lyrics on human emotions makes this analysis important.
\end{abstract}

%

%
\copyrightyear{2019} 
\acmYear{2019} 
\setcopyright{acmcopyright}
\acmConference[SIGIR '19]{Proceedings of the 42nd International ACM SIGIR Conference on Research and Development in Information Retrieval}{July 21--25, 2019}{Paris, France}
\acmBooktitle{Proceedings of the 42nd International ACM SIGIR Conference on Research and Development in Information Retrieval (SIGIR '19), July 21--25, 2019, Paris, France}
\acmPrice{15.00}
\acmDOI{10.1145/3331184.3331363}
\acmISBN{978-1-4503-6172-9/19/07}

%
\maketitle

\section{Introduction}
Music is an integral part of our culture. Smartphones and near ubiquitous availability of the internet have resulted in dramatic growth of online music consumption \cite{nielson18}. More than 85\% of online music subscribers search for song lyrics, indicating a keen interest of people in song lyrics. Song lyrics have a significant impact on human emotions and behavior. While listening to songs with violent lyrics can increase aggressive thoughts and hostile feelings \cite{Anderson}, listening to songs with pro-social lyrics increased the accessibility of pro-social thoughts, led to more interpersonal empathy, and fostered helping behavior \cite{Tobias}.

This paper is motivated by the observation that song lyrics have not received enough attention from the research community to understand them computationally. Several works focus on related problems such as personalized music recommendation \cite{Knees:2013:SMS:2559928.2542206}, song popularity prediction, and genre identification \cite{Fell2014LyricsbasedAA}.However, prior works on style analysis and bias measurement are limited only to the manual analysis of a few hundred songs. Such an approach cannot scale to the analysis of millions of songs. These works also fail to capitalize on recent advances in computational methods that generate a semantic representation for natural language text. Our work aims to fill this gap by applying these methods on a large corpus of song lyrics scraped from online user-generated content.

There are two main takeaway results from our work. First, popular songs significantly differ from other songs when it comes to the style of lyrics. This difference indicates that lyrics play a major role in deciding the popularity of a song. Second, biases in song lyrics correlate with biases measured in humans. This correlation indicates that song lyrics reflect the existing biases in society. To the best of our knowledge, this is the first work that analyzes song lyrics at the scale of half million lyrics to understand style and bias. Our results are reproducible as all our code and datasets are available publicly on the Web\footnote{https://github.com/manashpratim/Decoding-the-Style-and-Bias-of-Song-Lyrics}. We briefly review the related work in Section~\ref{secRelatedWork}. Our results are presented in Sections 3 and 4. Our conclusion and future work are highlighted in Section 5.

\section{Related Work}\label{secRelatedWork}
Song lyrics have been used for many tasks related to music mining such as genre identification and popularity prediction. Earlier works considered lyrics as a weak source of song characteristics as compared to auditory or social features. However, recent works have shown the strength of lyrics for music mining. Barman et al. have shown that knowledge encoded in lyrics can be utilized to improve the distributed representation of words \cite{manash}. Mayer et al. have introduced various features for lyrics processing \cite{mayer}. Fell and  Sporleder presented a lyrics-based analysis of songs based on vocabulary and song structure \cite{Fell2014LyricsbasedAA}. Our work complements these works by characterizing lyrics style using multiple attributes extracted from lyrics.

Many studies have analyzed gender and racial biases in song lyrics \cite{Eric, EJLLL49}. However, such an approach of manual analysis cannot scale to millions of songs. Caliskan et al. proposed Word Embedding Association Test (WEAT) to computationally measure biases in any text repository \cite{Car}. Their test quantifies biases by computing similarity scores between various sets of words. To compute similarity, the WEAT test represents words using a distributed word representation method such as fastText or word2vec \cite{NIPS2013_5021,joulin2017bag}. We apply the WEAT test on song lyrics and discuss its implications.


\begin{figure}[t]
    \centering
    \begin{subfigure}{0.22\textwidth}
    \centering
        \includegraphics[width=\textwidth]{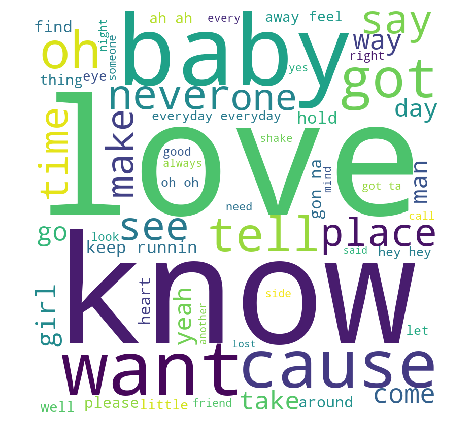}
        \caption{Top 100 Words in 1965}
        \label{fig:gull}
    \end{subfigure}\hfill
    \begin{subfigure}{0.22\textwidth}
    \centering
        \includegraphics[width=\textwidth]{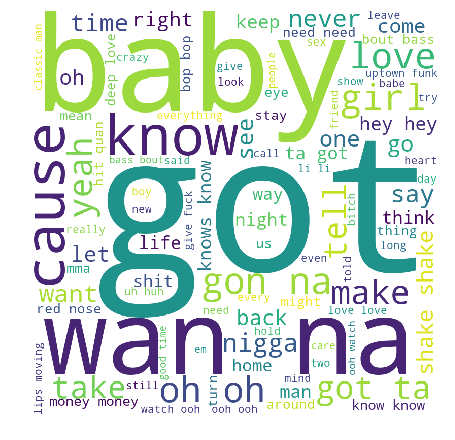}
        \caption{Top 100 Words in 2015}
        \label{fig:rat}
    \end{subfigure}
    ~ 
  
    \caption{Comparison of top words used}\label{fig1}
\end{figure}

\section{Style Analysis}
For style analysis, we created two datasets: Billboard (BB) and Million Song Dataset (MSD). For both the datasets, we obtained the song lyrics by scraping through the user-generated content on multiple websites such as MetroLyrics (\url{www.metrolyrics.com})and LyricsMode (\url{www.lyricsmode.com}). The BB dataset contains top 100 songs for each year (1965 to 2015) from the Billboard Hot 100 list (\url{www.billboard.com/charts/hot-100}). We consider these 5100 songs as popular songs. The original million song dataset only provides audio features and song metadata \cite{Bertin}. It does not provide song lyrics. Out of all songs listed in the original million song dataset, we could obtain lyrics only for 451,045 songs. We ensured that our BB and MSD datasets had no songs in common. Thus total songs in our analysis are around half a million. We had to do extensive data cleaning and preprocessing to use the scraped lyrics.

\begin{figure} [t]
	\centering
	\includegraphics[width=2.5in]{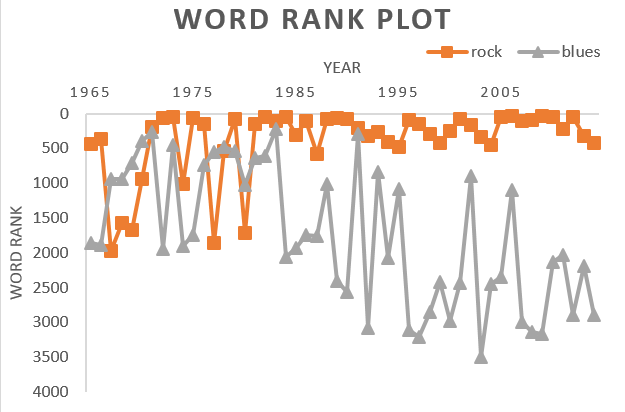}
	\caption{Year-wise rank comparison}
	\label{fig7}
\end{figure}

We started with an analysis of the vocabulary of song lyrics. Please refer to Figure~\ref{fig1}. This figure shows word clouds of the top 100 words used in popular song lyrics for years 1965 and 2015. We can observe that there is a major shift in vocabulary over time. For example, in 1965 the most popular word was ``love''. However, it was no more the case in 2015. To better visualize such trends, we have developed a word rank comparison tool\footnote{https://tiny.cc/songlyrics}. Given any set of words, this tool plots the relative popularity of those words in popular song lyrics through various years. Please refer to Figure~\ref{fig7}. This figure compares popularity of words ``rock" and ``blues" over the period from 1965 to 2015. For a given year $Y$, lower value of rank for a word $W$ indicates more frequent use of that word $W$ in popular song lyrics of year $Y$. We can observe that word ``rock" has maintained its popularity as compared to word ``blues".

\begin{figure}[b]
	\centering
	\includegraphics[width=2.5in]{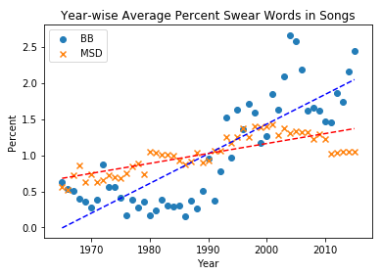}
	\caption{Comparison of swear word usage}
	\label{fig2}
\end{figure}

We also looked at the usage of swear words. To compile the list of swear words, we used various online resources. This list is available along with our code. Please refer to Figure~\ref{fig2}. This graph shows a comparison between popular (BB) and other (MSD) song lyrics based on swear word usage over the period from 1965 to 2015. We can observe that other songs have steadily gained the swear word usage over this period. From 1965 to 1995, popular songs used fewer swear words as compared to the other songs. However, from the 1990s there is a persistent trend of increasing swear word usage in popular songs. As compared to 1980s, swear words are used almost ten times more frequently now in popular song lyrics. Multiple studies have reported adverse effects of inappropriate content in music on the listeners \cite{hall}.

\begin{figure*}[bt]
    \centering
    \begin{subfigure}[b]{0.3\textwidth}
        \includegraphics[width=\textwidth]{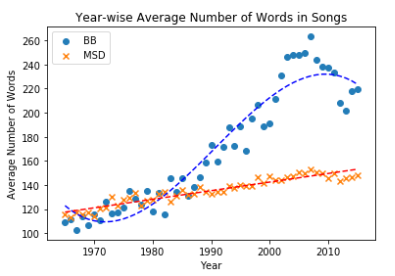}
        \caption{Year-wise Average Length}
        \label{fig:length}
    \end{subfigure}
    \qquad
    ~ 
    \begin{subfigure}[b]{0.3\textwidth}
        \includegraphics[width=\textwidth]{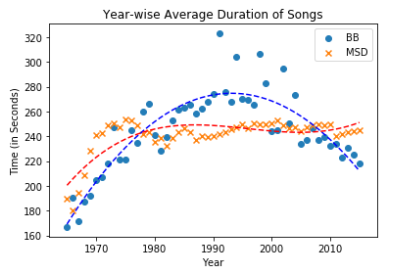}
        \caption{Year-wise Average Duration}
        \label{fig:duration}
    \end{subfigure}
    ~ 
     \qquad
    \begin{subfigure}[b]{0.3\textwidth}
        \includegraphics[width=\textwidth]{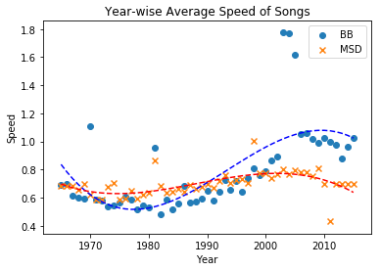}
        \caption{Year-wise Average Speed}
        \label{fig:speed}
    \end{subfigure}
    \caption{Comparison of Length, Duration, and Speed of songs}\label{fig3}
\end{figure*}

\begin{figure}[b]
	\centering
	\includegraphics[width=2.5in]{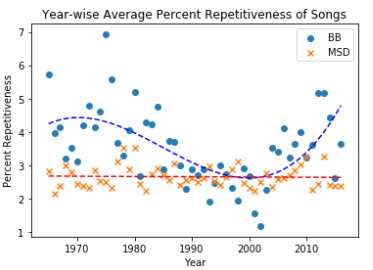}
	\caption{Comparison of repetitiveness}
	\label{fig:repeat}
\end{figure}

We also measured the length of song lyrics as the number of words in the song. Please refer to Figure~\ref{fig:length}. Other songs have shown a steady increase in length from 1965 to 2015. Popular songs also showed similar trend till 1980. However, since then popular songs are significantly more lengthy than other songs. Please refer to Figure~\ref{fig:duration}. This figure depicts the average duration of songs per year measured in seconds. Both poplar and other songs rose in duration till 1980. Since then, other songs have maintained the average duration of about 240 seconds. In contrast, popular songs were of longer duration from 1980 to 2010. However, the current trend shows far reduced duration for popular songs. Please refer to Figure~\ref{fig:speed}. This graph compares popular and other songs based on speed. We measure the speed of a song as (length in words/duration in seconds). We can observe that other songs have maintained an average speed of around 0.6 words per second. However, popular songs were comparatively slower till 1990 and since have become significantly faster than other songs. Some studies have reported that repetitive songs are lyrically processed more fluently and listeners prefer such songs \cite{usc}. We computed repetitiveness of a song lyric as ((1- (number of unique lines/total number of lines))*100). Please refer to Figure~\ref{fig:repeat}. Except for the period from 1990 to 2000, popular songs are more repetitive than other songs.

\begin{figure}[bt]
    \includegraphics[width=2.5 in]{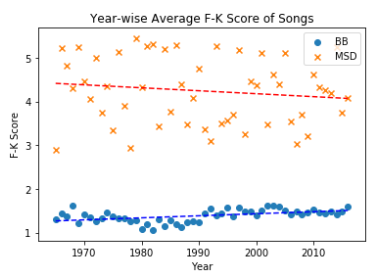}
    \caption{Comparison of FK score}
    \label{fk}
\end{figure}

Readability tests are standardized tests designed to provide a measure indicating how difficult it is to read or understand a given English text. We applied the well known Flesch Kincaid Readability Test (FK) on the song lyrics \cite{fscore}. The FK test returns a number that directly relates to the US Education grade system while also indicating the number of years of education required to understand the given English text. For example, an FK score of 5 indicates that anybody educated up to 5th grade can read and understand the given English text. Please refer to Figure~\ref{fk}. It can be seen that the FK scores of popular songs have always been less than 2. Also, the FK scores of other songs have always been quite higher than the popular songs.  This difference indicates that popular songs have always been easier to understand as compared to other songs.

\section{Bias Measurement}
We humans have certain biases in our thinking. For example, some people can find flower names more pleasant and insect names more unpleasant. These biases reflect in our various activities such as politics, movies, and song lyrics as well. Implicit Association Test (IAT) is a well-known test designed to measure such biases in human beings \cite{iat}. This test involves two sets of attribute words and two sets of target words. For example, consider two attribute word sets as pleasant words (nice, beautiful) and unpleasant words (dirty, awful). Also, consider two target word sets as flower names (rose, daffodil) and insect names (gnat, cockroach). The null hypothesis is that there should be little to no difference between the two sets of target words when we measure their similarity with the attribute word sets. The IAT test measures the unlikelihood of the null hypothesis by evaluating the effect size. A positive value of the IAT test indicates that people are biased to associate first attribute word set to the first target word set (Bias: flowers are pleasant) and second attribute word set with second target word set (Bias: insects are unpleasant). A negative value of the effect size indicates the bias in the other direction, that is flowers are unpleasant, and insects are pleasant. Larger magnitude of effect size indicates a stronger bias. If the value of effect size is closer to zero, then it indicates slight or no bias.

\begin{table*}[bt]
\centering
\caption{Comparison of Effect Size}
\label{table:algo}
\begin{tabular}{|c|c|c|c|c|c|c|}
\hline
  \bf Test no. & \bf Target Words & \bf Attribute Words &\bf w2v & \bf FT    &\bf CA   &\bf IAT    \\
\hline
1 & Flowers v/s Insects & Pleasant v/s Unpleasant & 1.04 & 1.34     & 1.5 & 1.35 \\
\hline
2 & Instruments v/s Weapons   & Pleasant v/s Unpleasant & 1.59 & 1.62  & 1.53 & 1.66    \\
\hline
3 & European-American v/s African-American names & Pleasant v/s Unpleasant & 0.71 & 0.6   & 1.41   & 1.17    \\
\hline
4 & Male v/s Female names  & Career v/s Family & 1.47 & 1.72   & 1.81   & 0.72    \\
\hline
5 & Math v/s Arts & Male v/s Female terms & 1.10 & 0.98 & 1.06 & 0.82 \\
\hline
6 & Science v/s Arts & Male v/s Female terms & 1.03 & 1.15  & 1.24 & 1.47        \\
\hline
7 & Mental v/s Physical disease & Temporary v/s Permanent & 0.89 & 1.15  & 1.38 & 1.01
\\
\hline
8 & Young v/s Old People's names & Pleasant v/s Unpleasant & 0.37 & 0.97 & 1.21 & 1.42
\\
\hline

\end{tabular}
\end{table*}

Caliskan et al. designed the Word Embedding Association Test (WEAT) by tweaking the IAT test \cite{Car}. Similar to the IAT test, this test can measure bias given the sets of attribute and target words. However, the IAT test requires human subjects to compute the bias value. On the other hand, the WEAT test can compute the bias value using a large text repository, and it does not require human subjects. The WEAT test represents attribute and target words as vectors using distributed representation methods such as word2vec and fastText \cite{NIPS2013_5021,joulin2017bag}. The WEAT test computes the similarity between words using the cosine similarity. Caliskan et al. have performed the bias measurement on a large internet crawl text corpus using the WEAT test. They have shown that their results correlate with the IAT tests conducted with human subjects.

We applied the WEAT test on our song lyrics dataset. Due to the small size of popular songs dataset, we cannot apply the WEAT test separately on popular songs lyrics. Please refer to Table~\ref{table:algo}. Corresponding to eight rows of the table, we have measured eight biases. We borrowed these attribute and target word sets from Caliskan et al. \cite{Car}. First two columns (w2v and FT) correspond to measurements on the song lyrics dataset with word2vec and fastText as the word embedding methods respectively. Next  column (CA) is the measurements reported by Caliskan et al. on a large internet crawl text corpus. The last column (IAT) is the original values reported by Greenwald et al. using human subjects \cite{iat}.

We can observe that for all eight tests, all four results columns have values with the same sign (positive values). This indicates that all four methods agree on the presence of the bias. However, they differ in the magnitude of the bias. We can consider the IAT column as the gold standard value because it is reported based on the human subjects. For six out of eight tests, FT values are closer to IAT than w2v values. This indicates superior performance of fastText over word2vec method to generate a semantic representation of words. For five tests out of eight, FT values are closer to IAT than CA values. For test number 3, 8, and 6 CA values are closer to IAT than FT. Test number 3 and 8 are based on various names. Our song lyrics dataset does not contain enough instances of these names. Therefore generated word vectors for these names are not accurate. Similarly, test 6 involves science related terms, and song lyrics dataset is deficient in such words. Our song lyrics dataset comprises of roughly 50 million tokes. This dataset is too small as compared to the dataset used by Caliskan et al. that contains over 840 billion tokes. Despite this huge difference in training data size, song lyrics are pretty good at capturing the human biases.

Out of all tests, we can see that the effect size of both FT and CA column is highest for test 4. This bias is about gender stereotypes and career paths. This bias can be expressed as males are more associated with career-oriented roles and females are more associated with family-oriented roles. This high effect value means that there is greater bias in these target groups. Doering and Th\'ebaud have shown that gender bias does not only disadvantage women, it could also disadvantage men \cite{laura}. Their findings showed that when gender stereotypes get attached to a role, it biases the authority that people attribute to the person who happens to work in that position. In a way, these stereotypes harm us all. With the dramatic growth in consumption of music \cite{nielson18}, such biases can reinforce the psychological status of these target groups \cite{Sapir}. Hence, it is crucial to address these prevalent biases in songs lyrics.

\section{Conclusion and Future Work}
We have analyzed over half a million lyrics to understand the style and prevalent biases. As compared to other songs, we have observed that popular songs have several distinguishing characteristics that can be expressed in terms of the style of lyrics. Lyrics can capture human biases quite accurately. This work can be extended further by investigating music genre-specific style and biases. 

\bibliographystyle{ACM-Reference-Format}
\bibliography{sample-base}

\end{document}